\documentclass[10pt,twocolumn]{ICSnewsletter}

\usepackage{hdg-packages}
\usepackage[url=true,backend=bibtex,maxnames=99,giveninits=true]{biblatex} 
\usepackage{hdg-commands}
\newcommand{\ecc}{\idx{ecc}}
\usepackage[]{fancyhdr}
\fancypagestyle{firstpage}
{
  \fancyhead[L]{\vspace{-2cm}cite this paper as\\
Herbert De~Gersem, Thomas Weiland, ''Reformulation and generalisation of the air-gap element'',\\ ICS Newsletter, Vol.~12, No.~1, ISSN~1026-0854, 1 March 2005, pp.~2--8, ArXiV.}
  \fancyhead[R]{}
}

\begin{document}

\title{Reformulation and Generalisation of the Air-Gap Element}
\maketitle

\thispagestyle{firstpage}

\begin{abstract}
The air-gap macro element is reformulated such that rotation,
rotor or stator skewing and rotor eccentricity can be incorporated
easily. The air-gap element is evaluated using Fast Fourier
Transforms which in combination with the Conjugate Gradient
algorithm leads to highly efficient and memory inexpensive
iterative solution scheme. The improved air-gap element features
beneficial approximation properties and is competitive to
moving-band and sliding-surface technique.
\end{abstract}

\section{Introduction}

In 1982, Abdel Razek et al. proposed the \emph{air-gap macro
element} to couple stator and rotor finite element (FE) models
taking their relative motion into account \cite{Abdel-Razek_1982aa}. In
this paper, it is shown that the air-gap element is not only
convenient for modelling rotation, but also for modelling skew and
eccentricity. Despite the clear advantages, the air-gap element
did not become standard in electrical machine simulation. This has
a practical, numerical reason: the air-gap element introduces a
cumbersome, dense block in the sparse FE system of equations. With
the rising popularity of iterative solvers as e.g. the Conjugate
Gradient algorithm, the air-gap element approach was abandoned in
favour of moving-band and sliding-surface techniques which both
guarantee sparse matrices.

In this paper, we want to rehabilitate the air-gap element
approach by showing its advantages over other techniques and by
alleviating its numerical drawbacks.


\section{FE Machine Model}

\begin{figure}[tb]
  \centering
  \includegraphics[width=4cm]{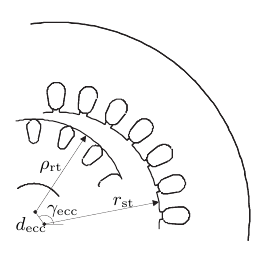}
  \caption{Eccentric geometry.}
  \label{fig:eccentric_geometry}
\end{figure}

For convenience, we only consider 2D FE models of cylindrical
machines. The generalisation of all discussed techniques to 3D
machine models is straightforward. Moreover, the benefits of the
presented techniques are even more pronounced for 3D models. We
assume the stator to be fixed and the rotor to be rotating at the
mechanical velocity $\omega\idx{m}(t)$. The center line of
rotation is not necessarily at the center of the stator
(Fig.~\ref{fig:eccentric_geometry}). Both center lines differ in
the considered 2D cross-section by the vector
$(d\ecc\cos\gamma\ecc,d\ecc\sin\gamma\ecc)$ where $d\ecc$ and
$\gamma\ecc$ denote the magnitude and the angle of the
eccentricity, respectively. Both static and dynamic eccentricities
are considered. Hence, $d\ecc$ and $\gamma\ecc$ may depend on
time. It is common to take the skewing of the stator or the rotor
into account, also when simulating 2D models. Here, we assume that
the rotor is skewed by an angle $\gamma\skw$. The standstill
cartesian and polar coordinate systems $(x,y)$ and $(r,\theta)$
are attached to the stator whereas the moving cartesian and polar
coordinate systems $(x',y')$ and $(\rho,\varphi)$ are considered
at the rotor. The relation between both coordinate systems can be
expressed by
\begin{equation}\label{eq:cdtransfo}
  r e^{\jmath\theta}=\rho e^{\jmath(\varphi+\omega\idx{m}t)}
  +d\ecc e^{\jmath\gamma\ecc} \;.
\end{equation}

\begin{figure}[t]
  \centering
  \includegraphics[width=8.1cm]{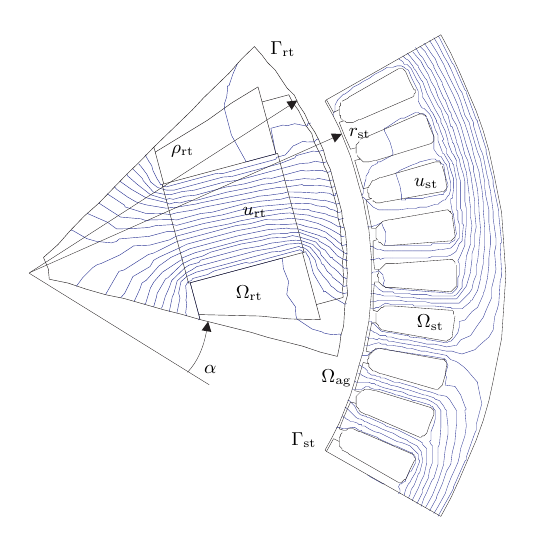}
  \caption{Computation domain $\Omega=\Omega\st\cup\Omega\ag\cup\Omega\rt$.}
  \label{fig:computational_domain}
\end{figure}

The computational domain
$\Omega=\Omega\st\cup\Omega\ag\cup\Omega\rt$ consists of the
stator part $\Omega\st$, the rotor part $\Omega\rt$ and the air
gap part $\Omega\ag$ (Fig.~\ref{fig:computational_domain}). The
interfaces between the stator and rotor parts and the air gap part
are denoted by $\Gamma\st=\Omega\st\cap\Omega\ag$ and
$\Gamma\rt=\Omega\rt\cap\Omega\ag$. The radii of both interfaces
are denoted by $r\st$ and $\rho\rt$ respectively. When the air gap
part is empty $\Omega\ag=\emptyset$, the air-gap element is
reduced to an interface condition applied at the common interface
$\Gamma\st=\Gamma\rt$ and weighted by harmonic functions
\cite{De-Gersem_2004ad}. The air-gap part $\Omega\ag$ of the
computational domain does not necessarily coincide with the
physical air gap, i.e., parts of the air gap may be comprised in
$\Omega\st$ and $\Omega\rt$.

A magnetoquasistatic formulation based on the magnetic vector
potential $\mbf{A}$ is discretised at the 2D cross-sections
$\Omega\st$ and $\Omega\rt$ by linear triangular finite elements
(FEs) yielding a system of equations of the form
\begin{equation}\label{eq:FEsystem}
  \mathbf{M}\frac{d}{dt}\mathbf{u}+\mathbf{K}\mathbf{u}+\mathbf{g}=\mathbf{f}
\end{equation}
for each FE model part. In (\ref{eq:FEsystem}), $\mathbf{M}$
denotes the mass matrix related to the conductivity, $\mathbf{K}$
is the curl-curl stiffness matrix incorporating the reluctivity,
$\mathbf{f}$ is the discretisation of the applied current density
and $\mathbf{u}$ contains the degrees of freedom for the
$z$-component of the magnetic vector potential. The contribution
\begin{equation}
  \mbf{g}_i = -\int_\Gamma H_\theta(\theta)N_i(\theta) \ud\Gamma
\end{equation}
represent the magnetic field strength $H_\theta$ tangential to the
interface $\Gamma$ weighted by the FE shape functions $N_i$. The
two disconnected FE domains $\Omega\st$ and $\Omega\rt$ give raise
to two decoupled FE systems of equations of the form
(\ref{eq:FEsystem}) which are here distinguished by subscripts
$\cdot\st$ and $\cdot\rt$. When $\Omega\ag=\emptyset$ and both FE
model parts have a matching mesh at the common interface, the
surface currents at the common boundary vanish, i.e.,
$\mbf{g}\st+\mbf{g}\rt=0$ and the standard FE system of equations
can be obtained by combining both FE systems and eliminating the
set of unknowns at one side of the interface. The time
discretisation of (\ref{eq:FEsystem}) can be carried out by any
time integration scheme.

The degrees of freedom for the $z$-component of the magnetic
vector potential allocated at $\Gamma\st$ and $\Gamma\rt$ are
collected in the vectors $\mbf{\tilde{u}}\st$ and
$\mbf{\tilde{u}}\rt$ and are selected from the vectors
$\mbf{u}\st$ and $\mbf{u}\rt$ by the rectangular selection
matrices $\mbf{Q}\st$ and $\mbf{Q}\rt$, i.e.,
$\mbf{\tilde{u}}\st=\mbf{Q}\st\mbf{u}\st$ and
$\mbf{\tilde{u}}\rt=\mbf{Q}\rt\mbf{u}\rt$.

\section{Stator-Rotor Coupling}


When classifying approaches for stator-rotor coupling according to
the nature of the geometrical decomposition of the machine model,
one can distinguish between \emph{sliding-surface} and
\emph{air-gap interface} models.

In \emph{sliding-surface} techniques, the movement of the rotor is
modelled at a common interface $\Gamma$ somewhere in the air gap.
For the \emph{locked-step} approach \cite{Preston_1988aa}, an
equidistant discretisation is applied at $\Gamma$ and the time
step is chosen such that the rotor rotates with an integral number
of mesh steps with respect to the stator. In general, the
restriction to matching meshes is too severe for technical machine
models. General sliding-surface techniques allow non-matching
grids at $\Gamma$ and differ in the kind of interpolation which is
applied at $\Gamma$, e.g. polynomial interpolation, mortar
projection \cite{Rodger_1990aa,Buffa_2000aa} or trigonometric
interpolation \cite{Demenko_1996aa,De-Gersem_2004ad}.


In \emph{air-gap interface} techniques, the air gap or a part of
it is considered as a separate domain $\Omega\ag$ with an
interface both to stator and rotor. The air-gap region is commonly
discretised by a technique which allows a more convenient
introduction of the relative motion than the standard FE method.
Examples are the boundary integral method \cite{Salon_1982aa},
the discontinuous Galerkin technique \cite{Alotto_2001ab}, the
air-gap macro element \cite{Abdel-Razek_1982aa} and a single-layer
moving band \cite{Davat_1985aa,Dular_2001ab}.

\section{Air-Gap Element}

The idea of the air-gap macro element is to model the air gap by a
truncated series of sines and cosines which converges to the
analytical solution if the number of FE vertices at $\Gamma\st$
and $\Gamma\rt$ tends to infinity
\cite{Abdel-Razek_1982aa,Lee_1991aa}. No additional degrees of
freedom are introduced. The air-gap element is traditionally
formulated as a particular transmission condition between the FE
vertices at $\Gamma\st$ and $\Gamma\rt$. The air-gap element idea
has been extended to higher-order elements \cite{Feliachi_1983aa} and
anti-periodic boundary conditions \cite{Flack_1994aa}. Attempts
to reduce the computational costs dedicated to the air-gap element
are reported in \cite{Flack_1994aa}.

This paper develops a general framework for the air-gap element
wherein the formulations of \cite{Abdel-Razek_1982aa},
\cite{Lee_1991aa} and \cite{Flack_1994aa} are comprised. The
paper frames the air-gap element within the more general theory of
spectral-element methods and offers an at first sight more
abstract but at the end more convenient notation. The more
efficient implementation is different from the approaches in
\cite{Abdel-Razek_1982aa}, \cite{Feliachi_1983aa} and \cite{Flack_1994aa}
and involves a so-called variational crime, i.e., the weakening of
the continuity conditions of the coupled discretisations. A major
concern is to keep the computational cost of the air-gap macro
element to a strict minimum.

The air-gap macro element is based on the analytical solution for
the 2D magnetic vector potential $A_z$ in the air gap of the 2D
cylindrical machine model
\begin{eqnarray}\label{eq:analytical}
  \nonumber\lefteqn{A_{{\rm ag},z}(r,\theta)
  =\mathbf{\hu{a}}_0+\mathbf{\hu{b}}_0\log\left(\frac{r}{\rho\rt}\right)} \\
    &&+\sum_{\lambda\in\Lambda}
    \left(\mathbf{\hu{a}}_\lambda \left(\frac{r}{\rho\rt}\right)^\lambda
    +\mathbf{\hu{b}}_\lambda \left(\frac{r}{\rho\rt}\right)^{-\lambda}
    \right)e^{-\jmath\lambda\theta}
\end{eqnarray}
where $\mathbf{\hu{a}}_\lambda$ and $\mathbf{\hu{b}}_\lambda$ are
complex-valued coefficients and $\lambda\in\Lambda$ is the
harmonic order. The air-gap element considers a finite set
$\Lambda$ of harmonic orders. In the formulation developed here,
the component with $\lambda=0$ is not considered. Instead, it is
assumed that a Dirichlet boundary condition is applied at both the
stator and rotor sides such that the arbitrary constant of $A_z$
is defined. In the case of a real-valued formulation,
$u(r,\theta)$ is real and hence $a_{-\lambda}=a_\lambda^*$ and
$b_{-\lambda}=b_\lambda^*$ where $^*$ stands for conjugate or
conjugate transpose for variables and matrices respectively.
Notice that by inserting Le Moivre's formula, i.e., $e^{-\lambda
p\lambda\theta}=\cos{p\lambda\theta}-\jmath\sin{p\lambda\theta}$,
the notation of \cite{Abdel-Razek_1982aa}, \cite{Feliachi_1983aa} and
\cite{Flack_1994aa} is regained. The magnetic field strength in
the $\theta$-direction is computed from (\ref{eq:analytical}):
\begin{eqnarray}\label{eq:Htheta}
  \nonumber\lefteqn{H_{{\rm ag},\theta}(r,\theta)=\mathbf{\hu{b}}_0\frac{\nu_0}{r}} \\
    &&+\sum_{\lambda\in\Lambda}
    \frac{\nu_0\lambda}{r}\left(\mathbf{\hu{a}}_\lambda \left(\frac{r}{\rho\rt}\right)^\lambda
    -\mathbf{\hu{b}}_\lambda
    \left(\frac{r}{\rho\rt}\right)^{-\lambda}:
    \right)e^{-\jmath\lambda\theta} \;.
\end{eqnarray}

The air-gap fields (\ref{eq:analytical}) and (\ref{eq:Htheta}),
evaluated at $\Gamma\st$ and $\Gamma\rt$, are matched to harmonic
decomposition at the interfaces, e.g.
\begin{eqnarray}
  \label{eq:Azst}
  A_{{\rm st},z} &=& \sum_{\lambda\in\Lambda}
    \mathbf{\hu{c}}_{{\rm st},\lambda}e^{-\jmath\lambda\theta} \;; \\
  \label{eq:Hthetast}
  H_{{\rm st},\theta} &=& \sum_{\lambda\in\Lambda}
    \mathbf{\hu{d}}_{{\rm st},\lambda}e^{-\jmath\lambda\theta} \;.
\end{eqnarray}
The coefficients $\mbf{\hu{c}}\st$ and $\mbf{\hu{c}}\rt$ are
related to $\mbf{\hu{a}}$ and $\mbf{\hu{b}}$ by
\begin{equation}\label{eq:T}
  \left[\begin{array}{c}
    \mbf{\hu{c}}\st \\ \mbf{\hu{c}}\rt
  \end{array}\right]
  =\mbf{T}
  \left[\begin{array}{c}
    \mbf{\hu{a}} \\ \mbf{\hu{b}}
  \end{array}\right]
\end{equation}
where $\mbf{T}$ is a block-diagonal matrix of which the blocks
$\mbf{T}_\lambda$ and $\mbf{T}_0$ are defined by
\begin{eqnarray}
  \mbf{T}_\lambda &=&
  \left[\begin{array}{cc}
    \xi^\lambda & \xi^{-\lambda} \\
    1 & 1
  \end{array}\right] \;; \\
  \mbf{T}_0 &=&
  \left[\begin{array}{cc}
    1 & \ln{\xi} \\
    1 & 0
  \end{array}\right]
\end{eqnarray}
and $\xi=r\st/\rho\rt$ denotes the air-gap radius ratio. The
coefficients $\mbf{\hu{d}}\st$ and $\mbf{\hu{d}}\rt$ are related
to $\mbf{\hu{a}}$ and $\mbf{\hu{b}}$ by
\begin{equation}\label{eq:G}
  \left[\begin{array}{c}
    \mbf{\hu{d}}\st \\ \mbf{\hu{d}}\rt
  \end{array}\right]
  =\mathbf{G}
  \left[\begin{array}{c}
    \mbf{\hu{a}} \\ \mbf{\hu{b}}
  \end{array}\right]
\end{equation}
where $\mbf{G}$ is a block-diagonal matrix of which the blocks
$\mbf{G}_\lambda$ are defined by
\begin{equation}
  \mbf{G}_\lambda =
  \nu_0\lambda\left[\begin{array}{cc}
    -\xi^\lambda/r\st & \xi^{-\lambda}/r\st \\
    -1/\rho\rt & 1/\rho\rt
  \end{array}\right]
\end{equation}
where $\nu_0$ denotes the reluctivity of air. The air-gap element
provides a so-called \emph{spectral Dirichlet-to-Neumann} operator
$\mathbf{G}\mathbf{T}^{-1}$
\begin{equation}
  \label{eq:SEsystem}
  \left[\begin{array}{c}
    \mathbf{\hu{d}}\st \\ \mathbf{\hu{d}}\rt
  \end{array}\right]
  = \mathbf{G}\mathbf{T}^{-1}
  \left[\begin{array}{c}
    \mathbf{\hu{c}}\st \\ \mathbf{\hu{c}}\rt
  \end{array}\right]
\end{equation}
mapping the harmonic coefficients for the magnetic vector
potential upon the corresponding harmonic coefficients for the
magnetic field strengths at $\Gamma\st$ and $\Gamma\rt$.

\section{Coupling with FE Model Parts}

The air-gap element is matched to the FE solutions at the
interfaces $\Gamma\st$ and $\Gamma\rt$. The necessary interface
conditions for $A_z$ and $H_\theta$ are applied in a \emph{weak
way}, i.e., by multiplying with test functions $w_p(\theta)$ and
$v_\zeta(\theta)$, e.g. at $\Gamma\st$
\begin{eqnarray}
  \label{eq:contAz}
  \int_{\Gamma\st} \left(A_{{\rm st},z}(r\st,\theta)
    -A_{{\rm ag},z}(r\st,\theta)\right) w_p(\theta) \ud\Gamma &=&0 \;; \\
  \label{eq:contHtheta}
  \int_{\Gamma\rt} \left(H_{{\rm st},\theta}(r\st,\theta)
    -H_{{\rm ag},\theta}(r\st,\theta)\right) v_\zeta(\theta) \ud\Gamma
    &=&0 \;.
\end{eqnarray}
In combination with appropriate test functions, (\ref{eq:contAz})
and (\ref{eq:contHtheta}) correspond to the mortar element
approach \cite{Buffa_2000aa}. The particular choice for test and
trial functions made here, however, corresponds to pointwise
matching at the vertices of the FE meshes:
\begin{eqnarray}
  w_p(\theta) &=& \delta_p(\theta) \;; \\
  v_\zeta(\theta) &=& e^{\jmath\zeta\theta} \;; \\
  h_q(\theta) &=& \delta_q(\theta) \;,
\end{eqnarray}
where the delta function $\delta_p(\theta)$ is defined such that
$\delta_p(\theta)$ is only non-zero at $\theta=\theta_p$ and its
integral over $\Gamma\st$ or $\Gamma\rt$ equals $2\pi r\st$ or
$2\pi\rho\rt$ respectively. The choices for $w_p(\theta)$ and
$h_q(\theta)$ in (\ref{eq:contAz}) and (\ref{eq:contHtheta}) are
such that the Fast Fourier Transform (FFT) algorithm can be used
for turning the FE degrees of freedom into the harmonic
coefficients and otherwise:
\begin{eqnarray}
  \label{eq:u2c}
  \left[\begin{array}{cc}
    \mathbf{\hu{c}}\st \\ \mathbf{\hu{c}}\rt
  \end{array}\right]
  &=& \left[\begin{array}{cc}
    \mathbf{F} & 0 \\ 0 & \mathbf{F}
  \end{array}\right]
  \left[\begin{array}{cc}
    \mathbf{\tilde{u}}\st \\ \mathbf{\tilde{u}}\rt
  \end{array}\right] \;; \\
  \label{eq:d2g}
  \left[\begin{array}{cc}
    \mathbf{\tilde{g}}\st \\ \mathbf{\tilde{g}}\rt
  \end{array}\right]
  &=& \left[\begin{array}{cc}
    \mathbf{F}^{-1} & 0 \\ 0 & \mathbf{F}^{-1}
  \end{array}\right]
  \left[\begin{array}{cc}
    \mathbf{\hu{d}}\st \\ \mathbf{\hu{d}}\rt
  \end{array}\right]
\end{eqnarray}
where $\mbf{F}$ denotes the FFT. Pointwise matching is a
\emph{collocation} approach for which it is known that the
discretisation error convergence less favourably as for the FE
method. In the following, matrices for the stator and rotor part
combined into block-diagonal matrices are denoted similarly, but
without subscript, i.e.,
\begin{equation}
  \mbf{K}=\left[\begin{array}{cc}
    \mbf{K}\st & 0 \\ 0 & \mbf{K}\rt
  \end{array}\right] \;.
\end{equation}

The combination of (\ref{eq:FEsystem}), (\ref{eq:SEsystem}),
(\ref{eq:u2c}) and (\ref{eq:d2g}) results in the coupled FE-SE
system of equations
\begin{equation}
  \label{eq:FESEsystem}
  \mathbf{M}\frac{d}{dt}\mathbf{u}
  +\mathbf{K}\mathbf{u}
  +\mathbf{Q}^T\mathbf{F}^{-1}\mathbf{G}\mathbf{T}^{-1}\mathbf{F}\mathbf{Q}\mathbf{u}
  =\mathbf{f} \;.
\end{equation}
Despite of the asymmetrical notation in (\ref{eq:FESEsystem}), it
is easy to prove that the \emph{air-gap stiffness matrix} part
\begin{equation}\label{eq:Kag}
  \mathbf{K}\ag=\mathbf{Q}^T\mathbf{F}^{-1}\mathbf{G}\mathbf{T}^{-1}\mathbf{F}\mathbf{Q}
\end{equation}
is a symmetric and real-valued operator. The symmetry follows from
the symmetry of the 2-by-2 diagonal blocks of
$\mathbf{G}\mathbf{T}^{-1}$ and the fact that
$\mathbf{F}^H=n\mathbf{F}^{-1}$ with $n$ the number of vertices at
the interfaces. Moreover, real-valued field distributions
$\mathbf{\tilde{u}}\st$ and $\mathbf{\tilde{u}}\rt$ are
transformed into harmonic components which satisfy
$\mathbf{\hu{c}}_{{\rm
st},-\lambda}=\overline{\mathbf{\hu{c}}_{{\rm st},\lambda}}$. This
property is maintained by the spectral Dirichlet-to-Neumann
operator $\mathbf{G}\mathbf{T}^{-1}$ from which can be concluded
that $\mathbf{K}\ag$ will return a real-valued vector.

\begin{figure}[tb]
  \centering\small
  (a)\includegraphics[width=3cm]{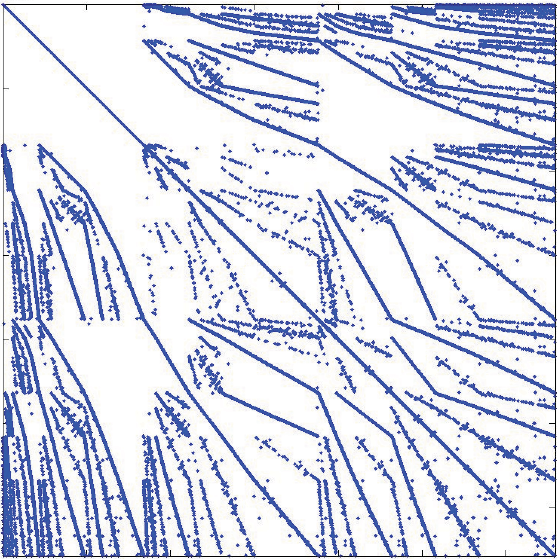} \hspace{0.5cm}
  (b)\includegraphics[width=3cm]{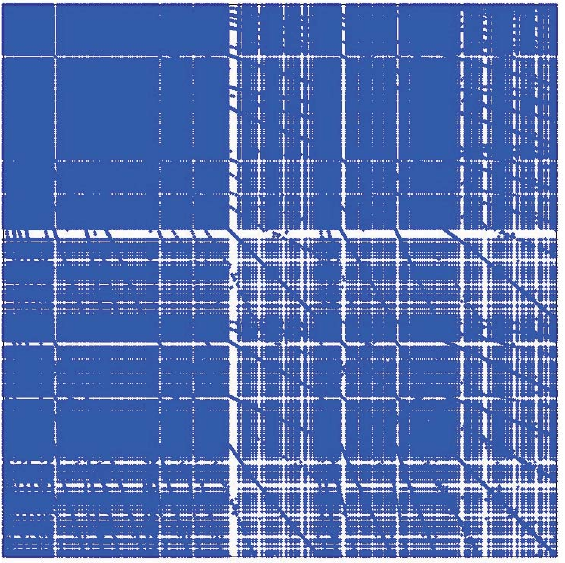}
    \caption{Sparsity patterns of (a) $\beta\mbf{M}+\mbf{K}$ (224172 nonzeros)
    and (b) $\mbf{A}=\beta\mbf{M}+\mbf{K}+\mbf{K}\ag$ (809040 nonzeros) where $\beta$
    is a factor determined by the time integration method.}
    \label{fig:spy}
\end{figure}

The air-gap stiffness matrix $\mbf{K}\ag$ contains a dense block
of dimension $2n\times 2n$. Hence, the insertion of $\mbf{K}\ag$
as an algebraic matrix into the FE system matrix would cause a
significant fill-in (Fig.~\ref{fig:spy}).
This is not a problem for a relatively small machine model, when
sufficient memory is available and direct system solution
techniques can be applied. For larger machine model, however, the
more expensive matrix-vector product leads to huge computation
times if standard iterative solution techniques are used.
Therefore, the explicit construction of $\mbf{K}\ag$ as an
algebraic matrix should be avoided by all means.

\section{Rotation}

When the rotor is rotated around its axis by an angle $\alpha$,
the new local coordinate system reads $(\rho,\varphi')$ where
$\varphi'=\varphi-\alpha$. The harmonic coefficients
$\mbf{\hu{c}}'\rt$ with respect to $(\rho,\varphi')$ are related
to $\mbf{\hu{c}}\rt$ by
$\mbf{\hu{c}}\rt=\mbf{R}_{-\alpha}\mbf{\hu{c}}'\rt$ where
$\mbf{R}_{-\alpha}$ is a diagonal matrix containing the phasors
\begin{equation}
  \mbf{R}_{-\alpha,\lambda,\lambda} =e^{\jmath\lambda\alpha} \;.
\end{equation}
This remarkably simple way to account for rotor displacement is
particularly advantageous in case of transient simulation. Using
this extended air-gap model, there is no longer need to rotate the
rotor mesh while time-stepping. It is sufficient to adapt the
\emph{rotation matrix} $\mbf{R}_{-\alpha}$ in the formulation to
the actual angular rotor position. The astonishing simplicity of
considering rotor motion is one of the major advantages of the
air-gap element \cite{De-Gersem_2004ad,De-Gersem_2005aa}.

\section{Rotor or Stator Skewing}

Rotor or stator skewing is commonly introduced in 2D machine
models by considering multiple slices at different axial positions
of the machine. The different slices are connected by the external
electric circuit \cite{Piriou_1990aa,Gyselinck_2001aa}. It is possible to
account for the skewing even if only 1 slice is modelled when the
spatial harmonic decomposition of the air-gap field is available
\cite{De-Gersem_2003aa}. The harmonic coefficients at one of the
interfaces, e.g. at $\Gamma\st$, are multiplied by analytical skew
factors depending on the skew angle $\gamma\skw$:
$\mbf{\hu{c}}'\st=\mbf{S}\skw\mbf{\hu{c}}\st$ where
\begin{equation}
  \mbf{S}_{{\rm skew},\lambda,\lambda}
  =\frac{2}{\lambda\gamma\skw}\sin\left(\frac{\lambda\gamma\skw}{2}\right)
  \;.
\end{equation}
Nevertheless, multiple slices have to be considered in order to
deal with the axial variation of the ferromagnetic saturation. In
that case, the skew angle of each slice is
$\gamma\skw\ell_{z,q}/\ell_z$ where $\ell_{z,q}$ and $\ell_z$ are
the axial length of slice $q$ and of the entire machine
respectively.

\section{Static and Dynamic Eccentricity}

In case of an eccentric rotor, the coordinate transformation
(\ref{eq:cdtransfo}) is introduced in the air-gap element formula
(\ref{eq:analytical}). The nested loop is rearranged into
increasing powers of $\rho$. It may be assumed that the magnitude
of the eccentricity is small compared to the radius of the rotor,
i.e., $\varepsilon=d\ecc/\rho\rt\ll 1$. Then, all higher-order
contributions $\varepsilon^k$ starting from $k=2$ can be neglected
in the series expansion:
\begin{eqnarray}
   \nonumber\lefteqn{A_z(\rho,\varphi)} && \\
   \nonumber && =\sum_{\lambda\in\Lambda}
   \Big[\left(\frac{\rho}{\rho\rt}\right)^\lambda\left(\mbf{\hu{a}}_\lambda
    +(\lambda+1)\varepsilon\ecc e^{-\jmath\gamma\ecc}\mbf{\hu{a}}_{\lambda+1}\right) \\
  \nonumber && +\left(\frac{\rho}{\rho\rt}\right)^{-\lambda}\left(\mbf{\hu{b}}_\lambda
    -(\lambda-1)\varepsilon\ecc e^{\jmath\gamma\ecc}\mbf{\hu{b}}_{\lambda-1}\right)
  \Big] e^{-\jmath\lambda\varphi} \\
  \label{eq:eccAz} && +\mathcal{O}\left(\varepsilon\ecc^2\right) \;.
\end{eqnarray}
The relative eccentricity and the eccentricity angle are combined
in a single complex number $\hu{\varepsilon}\ecc=\varepsilon\ecc
e^{\jmath\gamma\ecc}$. The azimuthal magnetic field strength with
respect to the rotor coordinate system $(\rho,\varphi)$ reads
\begin{eqnarray}
  \nonumber\lefteqn{H_\varphi(\rho,\varphi)} && \\
  \nonumber && =\sum_{\lambda\in\Lambda}
  \nonumber\frac{\nu_0\lambda}{\rho}\Big[
    \left(\frac{\rho}{\rho\rt}\right)^\lambda
    \left(\mbf{\hu{a}}_\lambda+(\lambda+1)\overline{\hu{\varepsilon}\ecc}\;\mbf{\hu{a}}_{\lambda+1}\right) \\
  \nonumber && -\left(\frac{\rho}{\rho\rt}\right)^{-\lambda}
    \left(\mbf{\hu{b}}_\lambda-(\lambda-1)\hu{\varepsilon}\ecc\;\mbf{\hu{b}}_{\lambda-1}\right)
  \Big]e^{-\jmath\lambda\varphi} \\
  \label{eq:eccHvarphi} && +\mathcal{O}\left(\varepsilon\ecc^2\right) \;.
\end{eqnarray}
The eccentricity is modelled as a perturbation of the classical
air-gap macro element \cite{De-Gersem_2006ac}. The
matrices $\mbf{T}_{\hu{\varepsilon}}$ and
$\mbf{G}_{\hu{\varepsilon}}$ are tri-diagonal block matrices where
the diagonal blocks are the same as in $\mbf{T}$ and $\mbf{G}$ and
the off-diagonal blocks are obtained by writing (\ref{eq:eccAz})
and (\ref{eq:eccHvarphi}) for $\rho=\rho\rt$ and collecting the
appropriate coefficients. An eccentric air-gap element is obtained
by replacing $\mbf{T}$ and $\mbf{G}$ in (\ref{eq:T}) and
(\ref{eq:G}) by $\mbf{T}_{\hu{\varepsilon}}$ and
$\mbf{G}_{\hu{\varepsilon}}$.

\section{Arbitrary FE Meshes at $\Gamma\st$ and $\Gamma\rt$}

The assumption that the FE meshes at $\Gamma\st$ and $\Gamma\rt$
should be equidistant and should have the same number of nodes, is
too restrictive for technically relevant models. The classical FFT
algorithm, however, dictates the uniform distribution of FE
vertices at $\Gamma\st$ and $\Gamma\rt$. Recently, FFT algorithms
for non-equispaced data are developed \cite{Potts_2001aa}. Their
application is slightly more expensive compared to the classical
FFT algorithm. The computational complexity, however, remains
$\mathcal{O}(n\log n)$ such that the overall computation time of the
FE simulation is barely influenced by the application of the
air-gap element. When different numbers of FE vertices are applied
at $\Gamma\st$ and $\Gamma\rt$, the sets of harmonic orders
present in $\mbf{\hu{c}}\st=\mbf{F}\st\tilde{\mbf{u}}\st$ and
$\mbf{\hu{c}}\rt=\mbf{F}\rt\tilde{\mbf{u}}\rt$, denoted by
$\Lambda\st$ and $\Lambda\rt$ respectively, are different. The
coupling through the eccentric air-gap element is only applied for
a user-defined subset of the available components:
$\Lambda\subset\Lambda\st\cup\Lambda\rt$. The selectors
$\mbf{X}\st$ and $\mbf{X}\rt$ select the harmonic coefficients
with orders $\lambda\in\Lambda$ from $\mbf{\hu{c}}\st$ and
$\mbf{\hu{c}}\rt$ respectively. Only these components are treated
by the modified air-gap reluctance matrix:
\begin{equation}
  \mbf{K}\ag
  =\mbf{Q}^T\mbf{F}^{-1}\mbf{X}^T\mbf{G}
    \mbf{T}^{-1}\mbf{X}\mbf{F}\mbf{Q} \;.
\end{equation}
The harmonic coefficients which are not considered by the air-gap
element
are not influenced by the air-gap and experience homogeneous
Neumann boundary conditions at $\Gamma\st$ and $\Gamma\rt$.

\section{Procedure}

\begin{figure*}[tb]
  \centering\small
  \includegraphics[width=13cm]{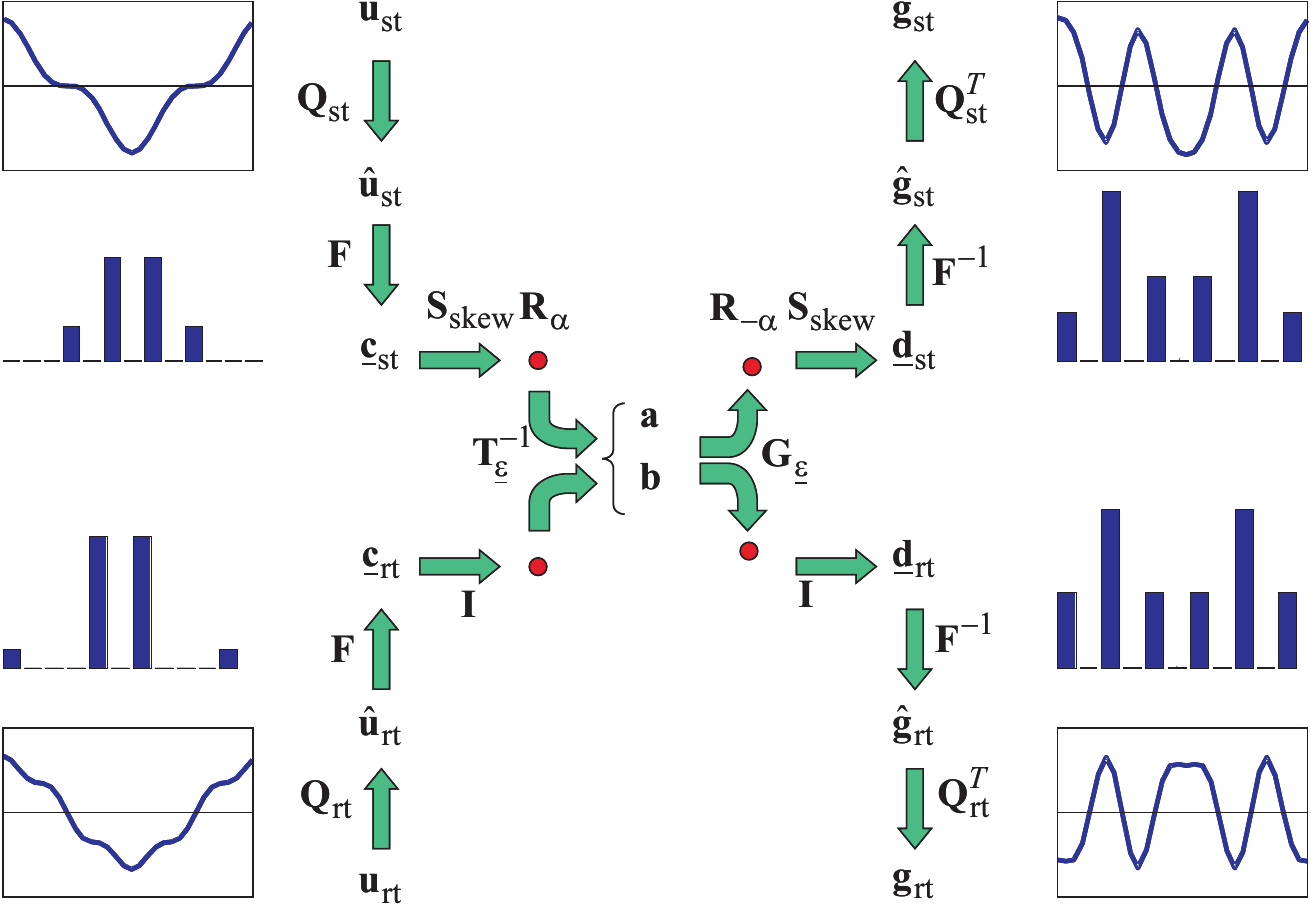}
  \caption{Procedure to apply the eccentric air-gap element
  coupling the FE model parts of stator and rotor.}
  \label{fig:routine}
\end{figure*}

After introduction of rotation, skewing and eccentricity, the
air-gap stiffness matrix reads
\begin{equation}
  \mbf{K}\ag
  =\mbf{Q}^T\mbf{F}^{-1}\mbf{X}^T\mbf{R}_\alpha\mbf{S}\skw\mbf{G}_{\hu{\varepsilon}}
    \mbf{T}_{\hu{\varepsilon}}^{-1}\mbf{S}\skw\mbf{R}_{-\alpha}\mbf{X}\mbf{F}\mbf{Q} \;.
\end{equation}
The heavy matrix notation used for developing the air-gap element
in the previous sections is misleading. The application of the
air-gap element as a \emph{procedure} acting upon the
distributions of the magnetic vector potential at $\Gamma\st$ and
$\Gamma\rt$ provides both a better understanding and a more
efficient algorithm. The air-gap element computes the surface
currents $\mbf{\tilde{g}}\st$ and $\mbf{\tilde{g}}\rt$ occurring
at $\Gamma\st$ and $\Gamma\rt$ for particular field distributions
$\mbf{\tilde{u}}\st$ and $\mbf{\tilde{u}}\rt$ applied at the same
boundaries. The procedure for explicitly carrying out this
computation is depicted in Fig.~\ref{fig:routine}.
\begin{itemize}
\item Given a temporary FE solution vector $[\mbf{u}\st\; \mbf{u}\rt]^T$ for
the magnetic vector potentials in $\Omega\st$ and $\Omega\rt$, the
values at the air-gap interfaces $\Gamma\st$ and $\Gamma\rt$ are
selected by the selection matrices $\mbf{Q}\st$ and $\mbf{Q}\rt$,
yielding $\mbf{\tilde{u}}\st$ and $\mbf{\tilde{u}}\rt$. This
operations are implemented on a computer by subscripting the FE
solution vectors by predefined index sets.
\item The vectors of harmonics $\mbf{\hu{c}}\st$
and $\mbf{\hu{c}}\rt$ are obtained by applying Fast Fourier
Transforms to $\mbf{\tilde{u}}\st$ and $\mbf{\tilde{u}}\rt$. The
FFTW algorithm which is used for that purpose, consists of a setup
step, which is only carried out once, and a computation step which
has to be invoked 4 times per application of $\mbf{K}\ag$
\cite{Frigo_1998aa}.
\item At one of both sides, the rotation operator $\mbf{R}_{-\alpha}$ and the skewing
operator $\mbf{S}\skw$ are applied. This corresponds to a
vector-vector multiplication by phasors and scaling factors
respectively.
\item From the (modified) harmonic coefficients $\mbf{\hu{c}}\st$ and
$\mbf{\hu{c}}\rt$ related to the magnetic vector potential
distribution at $\Gamma\st$ and $\Gamma\rt$, the harmonic
coefficients $\mbf{\hu{d}}\st$ and $\mbf{\hu{d}}\rt$ related to
the fictitious surface currents at $\Gamma\st$ and $\Gamma\rt$ are
computed by applying the precomputed 2-by-2 blocks
$\mbf{G}_\lambda\mbf{T}_\lambda^{-1}$ or in the eccentric case by
applying
$\mbf{G}_{\hu{\varepsilon}}\mbf{T}_{\hu{\varepsilon}}^{-1}$.
\item The discretized surface currents $\mbf{\tilde{g}}\st$ and $\mbf{\tilde{g}}\rt$
at $\Gamma\st$ and $\Gamma\rt$ follow by an inverse Fourier
transformation applied to $\mbf{\hu{d}}\st$ and $\mbf{\hu{d}}\rt$.
\item Finally, $\mbf{\tilde{u}}\st$ and $\mbf{\tilde{u}}\rt$ are prolongated
to vectors according to the FE problems by computing
$\mbf{g}\st=\mbf{Q}\st^T\mbf{\tilde{g}}\st$ and
$\mbf{g}\rt=\mbf{Q}\rt^T\mbf{\tilde{g}}\rt$ again by invoking
subscripting using appropriate index sets on the computer.
\end{itemize}
This procedure is made available as a separate software routine
\begin{equation}\label{eq:routine}
  \mathtt{g=airgapelement(u,alpha,gammaskew,decc,gecc)}
\end{equation}
taking the vectors $\mbf{u}\st$ and $\mbf{u}\rt$, the rotation
angle $\alpha(t)$, the skew angle $\gamma\skw$ and the magnitude
$d\ecc(t)$ and angle $\gamma\ecc(t)$ of the eccentricity as
parameters and returning the vectors $\mbf{g}\st$ and
$\mbf{g}\rt$.

%

\section{Discretisation and Consistency Errors}

\subsection{Discretisation error of the air-gap element}

When the range of the harmonic orders $\lambda$ is limited to
$\Lambda$ instead of $\lambda\in\{,-\infty,\ldots,+\infty\}$ as
for the analytical model, a truncation error is introduced. The
analytical air-gap solution is equivalent to a spectral element
(SE) discretization using an orthogonal set of harmonic functions
as test and trial functions \cite{Fornberg_1996aa}. The SE approach is
known to achieve an exponential convergence rate of the
discretization error as long as no sharp corners and no jumps of
the material coefficients are present in the SE model part, which
is the case for the air gap of an electrical machine. Hence, the
overall discretization error of the coupled FE-SE formulation will
be dominated by the discretization error of the FE model parts and
the consistency error introduced by the non-matching
discretizations at $\Gamma\st$ and $\Gamma\rt$.

\subsection{Consistency error at the interfaces}

A consistency error is introduced by the choice of non-matching
shape functions for $A_z$ and $H_\theta$ at the interfaces. It is
shown in literature dealing with the mortar-element method (see
e.g. \cite{Rapetti_2000aa}) that the consistency error at
$\Gamma\st$ and $\Gamma\rt$ may have a lower convergence rate than
the one for the FE and SE discretizations at $\Omega\st$,
$\Omega\rt$ and $\Omega\ag$, which leads to a degenerated
convergence of the overall discretization error for the hybrid
model. Moreover, for triangulations of cylindrical machines,
$\Gamma\st$ and $\Gamma\rt$ are polygons instead of circles such
that the hybrid integrals in (\ref{eq:contAz}) and
(\ref{eq:contHtheta}) connect the SE degrees of freedom not only
to the FE degrees of freedom at the nodes of the interfaces but
also to the FE degrees of freedom associated with the first layer
of nodes inside $\Omega\st$ and $\Omega\rt$ \cite{Rapetti_2000aa}.
The approach in  \cite{Rapetti_2000aa} offers a better convergence
of the consistency error but causes a substantial increase of the
computational work, especially because FFT is not longer
applicable. An acceptable trade-off consists of integrating the
interface conditions at the circles $r=r\st$ and $\rho=\rho\rt$
and assuming a linear variation of the FE shape functions at the
circle segments. This improvement is introduced in the formulation
described above by replacing $\mathbf{F}$ by $\mathbf{S}_{\rm int}
\mathbf{F}$ where
\begin{equation}
  \mathbf{S}_{{\rm int},\lambda,\lambda}=\frac{2n}{\lambda}\sin(\frac{\lambda}{2n})\;.
\end{equation}
This approach exactly corresponds to the original macro element
\cite{Abdel-Razek_1982aa}. By substituting
$\cos{\lambda\theta}-\jmath\sin{\lambda\theta}$ for
$e^{-\jmath\lambda\theta}$, the notation used in
\cite{Abdel-Razek_1982aa} and \cite{Flack_1994aa} is achieved.

\subsection{Stability}

\begin{figure}[tb]
  \centering
  \includegraphics[width=8cm]{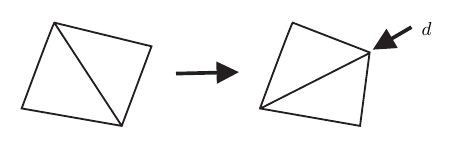}
  \caption{Topological change of the FE mesh due to a small
  displacement of a part of the model. The displacement $d$
  of one node belonging to the rotor causes a
  small distortion of a quadrilateral consisting of two triangles in the moving band.
  The diagonal edge of the original mesh is now shorter than the edge between the two
  other opposite nodes. In that case, the Delaunay
  mesh procedure selects the other diagonal as a division of the
  quadrangle into two triangles and thereby changes the topology of the mesh.}
  \label{fig:displacement}
\end{figure}

For a \emph{stable} numerical approach, arbitrarily small
differences in the results can be uniformly bounded by arbitrarily
small differences in the input data. When simulating solid-body
motion, the modelling of the geometrical changes is an important
part of the simulation scheme. When using a moving-band technique,
the moving band is remeshed between two successive simulation
steps. An arbitrary small displacement of the rotor (either
rotation or eccentricity) can result in a topological change of
the FE mesh, e.g. a different number of FE vertices or an edge
that is oriented otherwise (Fig.~\ref{fig:displacement}).
Hence, the remeshing of the moving band is not stable with respect
to the solid-body displacement of the rotor, i.e., there does not
exist a constant $C<\infty$ such that
\begin{equation}
  \left\| \mbf{u}(d)-\mbf{u}(0) \right \| < C \| d \|, \quad\forall \| d \|>0
\end{equation}
where $\mbf{u}(d)$ represent the FE solution for a solid-body
displacement $d$. Such changes have a significant influence upon
the computed FE solutions. As a consequence, it is not always
possible to obtain an arbitrary small difference between two
successive FE solutions even if a very small displacement $d$ is
applied. A similar phenomenon is reported as \emph{torque ripple}
when simulating rotating machines using the moving-band approach
\cite{Tsukerman_1995ab}. The moving-band technique has been
generalized for FE machine models with static and dynamic
eccentricity \cite{De-Bortoli_1993aa,Tenhunen_2001ab}. Although a
moving-band approach which is stable with respect to eccentricity
has been developed \cite{Schlensok_2004aa}, the stability with
respect to rotation is still not given. This phenomenon is
particularly annoying when the displacements are smaller than the
FE mesh size, as is commonly the case when accounting for
eccentricity in FE machine models. One of the major benefits of
the eccentric air-gap element, is the guaranteed stability of the
air-gap discretization with respect to small displacements of the
rotor. This property is indispensable to ensure a reliable torque
and force computation.

\section{Iterative Solution}

The assembly of $\mathbf{K}\ag$ into $\beta\mathbf{M}+\mathbf{K}$
drastically increases the density of the matrix and, as a
consequence, will decrease the numerical efficiency of the air-gap
element approach. To our opinion, this is the major reason why the
air-gap element approach is not so widespread as the moving-band
technique. Here, an iterative approach is proposed which
alleviates this problem. The key point is that the air-gap
stiffness matrix is not assembled into the FE system matrix.
Instead, the air-gap stiffness matrix $\mbf{K}\ag$ is provided as
a routine (\ref{eq:routine}) turning an arbitrary vector $\mbf{u}$
of FE degrees of freedom into the corresponding surface currents
$\mbf{g}$ reflecting the reluctance of the air gap as illustrated
by Fig.~\ref{fig:routine}. At the $k$-th CG step, the
matrix-vector product $\mbf{f}_k=\mbf{A}\mbf{u}_k$ is carried out
by adding the contribution of the conventional matrix-vector
product of $\mbf{u}_k$ by $\beta\mbf{M}+\mbf{K}$ and the vector
$\mbf{g}_k$ obtained by applying the routine (\ref{eq:routine}) to
$\mbf{u}_k$. The operations in (\ref{eq:Kag}) needed to compute
$\mbf{g}_k$ from $\mbf{u}_k$ are applied successively as depicted
in Fig.~\ref{fig:routine}. By applying the Fast Fourier Transform
(FFT) algorithm for $\mbf{F}$ and $\mbf{F}^{-1}$, the
computational cost of the discrete Fourier transforms, and thus of
the eccentric air-gap element, can be kept as low as
$\mathcal{O}(n\log{n})$. The application of FFT is absolutely
necessary to ensure that the eccentric air-gap element is
competitive to moving-band and sliding-surface approaches
\cite{De-Gersem_2005aa}. Inserting the operator (\ref{eq:Kag}) and
combining real-valued and complex-valued arithmetic is not
possible in many black-box CG solvers. Hence, a specialized CG
algorithm has to be developed, which is the major disadvantage of
the improved air-gap element approach described here.

The convergence of CG can be improved considerably by applying
preconditioning. Since no algebraic matrix representing $\mbf{A}$
is available, pure algebraic preconditioning techniques such as
e.g. Successive Overrelaxation, Incomplete Cholesky and Algebraic
Multigrid (AMG) are not applicable \cite{De-Gersem_2005aa}. A
straightforward preconditioner is e.g. an additive Schwarz
approach
\begin{equation}
  (\mbf{I}-\mbf{Q})^T \tilde{\mbf{A}}^{-1} (\mbf{I}-\mbf{Q})
  +\mbf{Q}^T\mbf{F}^{-1}\mbf{T}_{\hu{\varepsilon}}\mbf{G}_{\hu{\varepsilon}}^{-1}\mbf{F}\mbf{Q}
\end{equation}
where $\tilde{\mbf{A}}^{-1}$ denotes the application of 1 V-cycle
of AMG. Also the preconditioning step involves operations defined
by a routine similar to (\ref{eq:routine}).

Special care has to be taken if the constant components
$\mathbf{\hu{a}}_0+\mathbf{\hu{b}}_0\log(r/\rho\rt)$ are
considered in the air-gap element. The solution for $A_z$ is known
up to a constant field. When e.g. only Dirichlet constraints are
applied to the stator model part, $\mathbf{K}\rt$ is singular and
the constant at the rotor side should be fixed by the air-gap
element formulation. Then, a problem rises for the magnetic field
strength because $\mathbf{G}_0$ is a singular matrix.
$\mathbf{K}\ag^{-1}$ is not defined for the constant components
such that the analytical formulae (\ref{eq:analytical}) and
(\ref{eq:Htheta}) fail to compute $\mathbf{\hu{d}}_{{\rm st},0}$
and $\mathbf{\hu{d}}_{{\rm rt},0}$ from $\mathbf{\hu{c}}_{{\rm
st},0}$ and $\mathbf{\hu{c}}_{{\rm rt},0}$. The constant component
of the magnetic field strength is determined by the total current
$i\rt$ through the rotor:
\begin{eqnarray}
  H_{{\rm st},\theta} &=& \frac{i\rt}{2\pi r\st} \\
  H_{{\rm rt},\theta} &=& \frac{i\rt}{2\pi \rho\rt}
\end{eqnarray}
which necessitates the coupling of the air-gap element to the
degrees of freedom of the external circuit model. An easier
procedure consists of applying at least one Dirichlet boundary
condition at the rotor side and replacing the 2-by-2 block
matrices $\mathbf{G}_0\mathbf{T}_0^{-1}$ and
$\mathbf{T}_0\mathbf{G}_0^{-1}$ by zero matrices such that no
constant components have to be propagated through the air-gap
model. This approach is, however, not applicable for machine
models where the total current through the rotor does not vanish.

\section{Torque and Unbalanced Magnetic Pull}

When the air-gap harmonics are available, a highly accurate value
for the torque can be computed using the formula
\cite{Mertens_1997aa}
\begin{equation}\label{eq:torque}
  T=\sum_{\lambda\in\Lambda} 8\pi\nu_0\ell_z{\rm Im}
    \left\{\mathbf{\hu{a}}^H\mathbf{\hu{b}}\right\} \;.
\end{equation}
Thanks to the stability of the air-gap element approach, torque
ripple is avoided \cite{Demenko_1996aa}.

The availability of the harmonic coefficients of the air-gap field
also allows to compute the unbalanced magnetic pull up to the
highest possible accuracy achieved by the FE model. The magnetic
flux density $(B_r,B_\theta)$ with respect to the standstill polar
coordinate system
can be gathered into a complex-valued field:
\begin{equation}\label{eq:Bcpl}
  \hu{B} =B_r+\jmath B_\theta
   =\sum_{\lambda\in\Lambda}
  \frac{-\jmath\lambda}{r}2\mbf{\hu{a}}_\lambda
  \left(\frac{r}{\rho\rt}\right)^\lambda e^{-\jmath\lambda\theta}
  \;.
\end{equation}
Similarly, the force components $F_x$ and $F_y$, expressed by the
Maxwell stress tensor, can be brought together:
\begin{equation}
  \label{eq:Fcpl}\hu{F} =F_x+\jmath F_y
  =\ell_z\int_0^{2\pi} \frac{\nu_0}{2}\hu{B}^2
  e^{\jmath\theta}r\ud\theta \;.
\end{equation}
Introducing (\ref{eq:Bcpl}) into (\ref{eq:Fcpl}) and working out
the integral leads to
\begin{equation}
  \underline{F}=-\frac{4\pi\ell_z\nu_0}{\rho\rt}
  \sum_{\lambda\in\Lambda}
  \lambda(1-\lambda)\mbf{\hu{a}}_\lambda\mbf{\hu{a}}_{1-\lambda} \;.
\end{equation}

\section{Example}

\begin{figure*}[tb]
  \centering\small
  (a)\includegraphics[width=4.1cm]{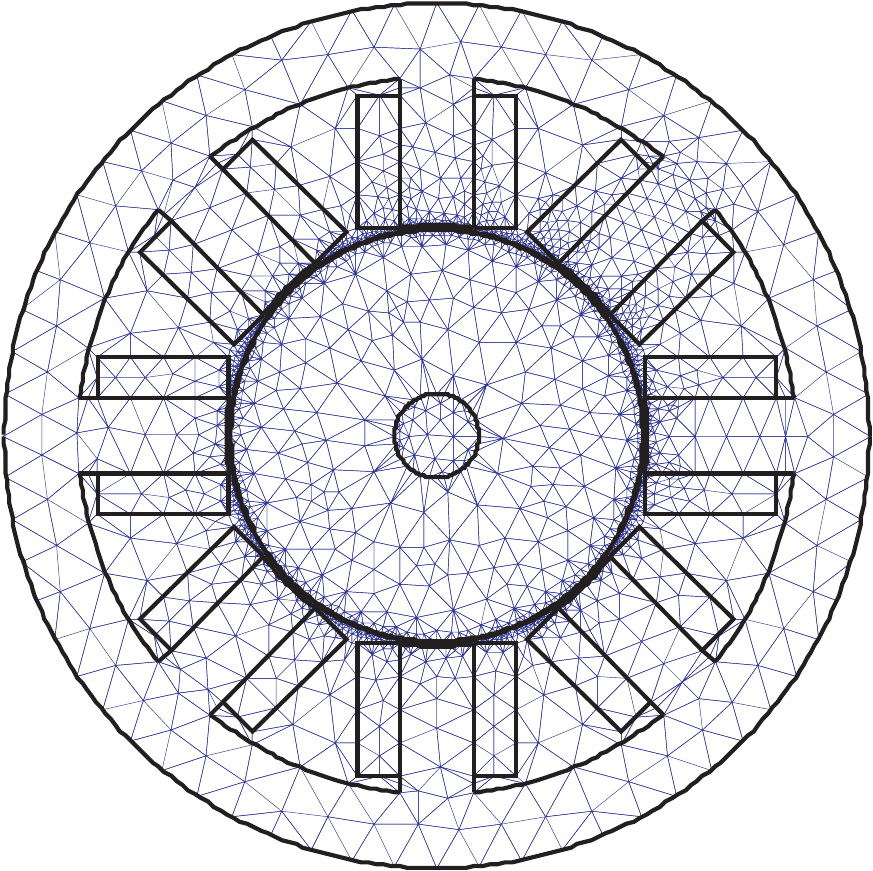} \hspace{1cm}
  (b)\includegraphics[width=4.1cm]{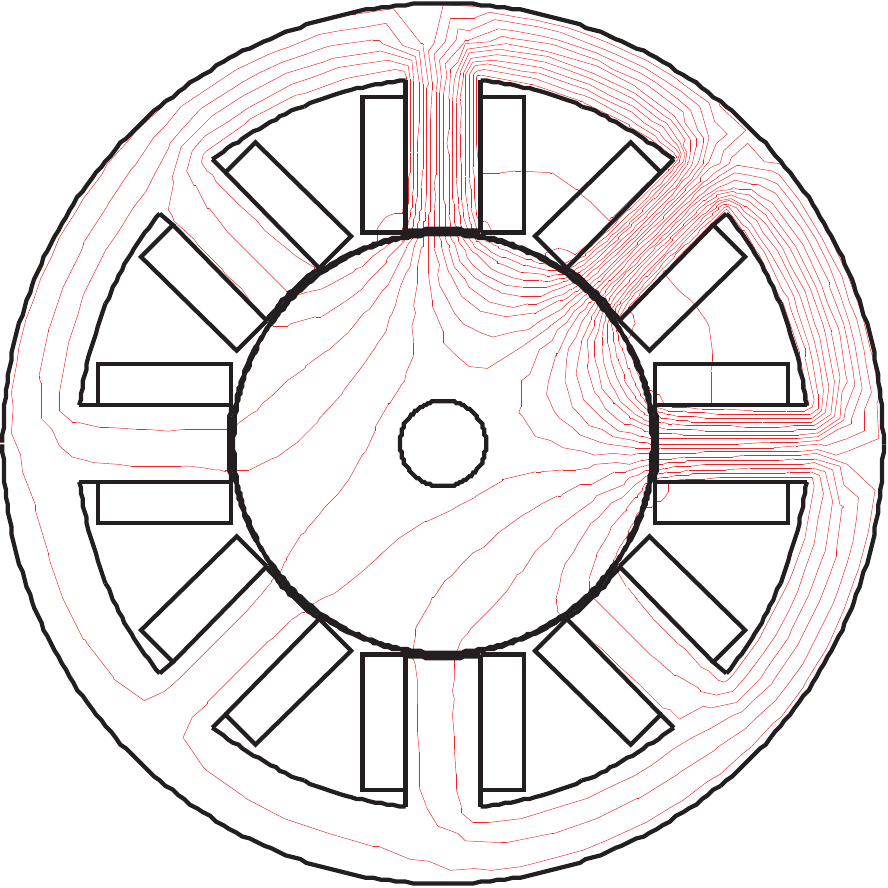} \hspace{1cm}
  (c)\includegraphics[width=4.1cm]{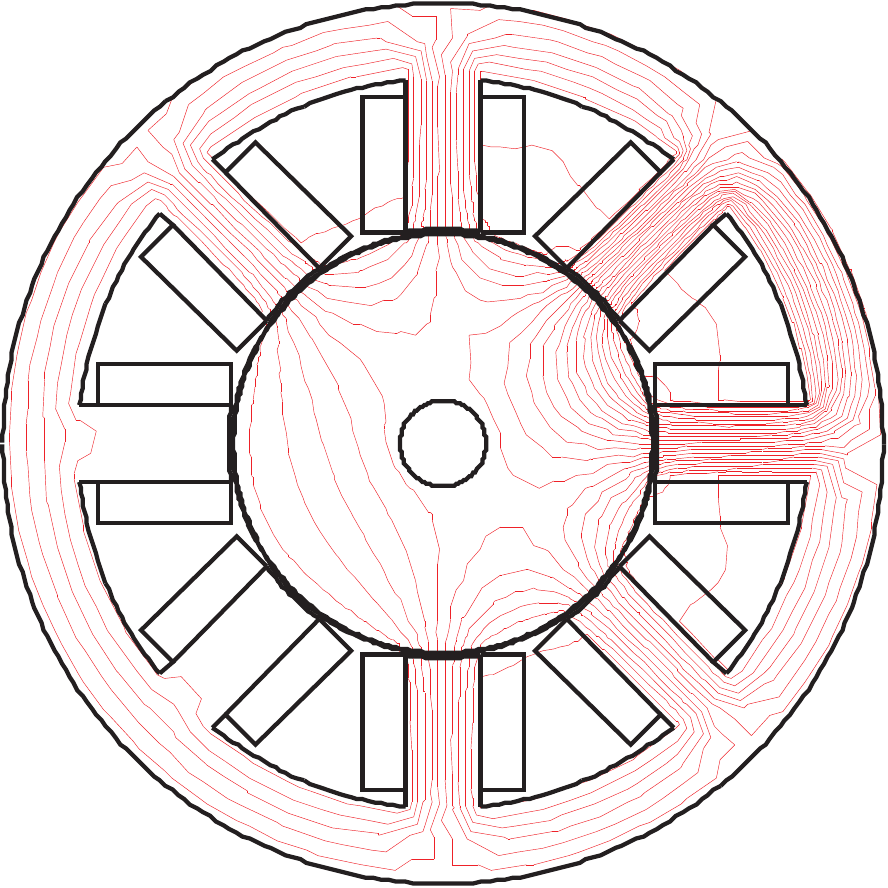}
  \caption{Magnetic-bearing model: (a) FE mesh;
  (b) magnetic flux lines (unbiased excitation) and
  (c) magnetic flux lines (biased excitation).}
  \label{fig:alex}
\end{figure*}

The 2D FE model of a magnetic bearing (Fig.~\ref{fig:alex}a) is
equipped with the eccentric air-gap element.
Transient simulations are carried out to test the numerical
behaviour of the eccentric air-gap element and the embedded force
computation. The rotor is submitted to a prescribed movement which
is a combination of a translation from the position
$(0.5\;\mathrm{mm},9\pi/8)$ to the position
$(0.5\;\mathrm{mm},\pi/8)$ and a rotation around its axis. The
stator coils are excited such that a force under an angle of
$22.5$ degrees is generated. Both unbiased and biased current
excitations are considered. The magnetic fluxes when the rotor is
at the center position are shown for both unbiased and biased
excitation in Fig.~\ref{fig:alex}.

During the transient simulation, the stator and rotor FE meshes do
not have to be reconstructed. The movement of the rotor between
two successive time steps only affect the operators
$\mbf{T}_{\hu{\varepsilon}}$, $\mbf{G}_{\hu{\varepsilon}}$ and
$\mbf{R}_\alpha$ which are embedded in the eccentric air-gap
stiffness operator $\mbf{K}\ag$. In practice, only the parameters
specified in the routine (\ref{eq:routine}) have to be adapted.
The FE systems $\beta\mbf{M}\st+\mbf{K}\st$ and
$\beta\mbf{M}\rt+\mbf{K}\rt$ are only reassembled as to account
for the saturation of the ferromagnetic parts. The AMG
preconditioner is only constructed once before starting the
time-stepping procedure. The formulation based on the eccentric
air-gap element substantially diminishes the time for system
set-up during the transient simulation. No ripple on the results
for the force is encountered, indicating the stability of the
discretization scheme with respect to small displacements.

\section{Conclusions}

The reformulation of the air-gap element and its interpretation as
a spectral-element discretisation in the air gap leads to a
formulation for which the major part of the work can be carried
out by Fast Fourier Transforms. Rotor rotation, stator or rotor
skewing and rotor eccentricity can be incorporated in the air-gap
element in a natural and convenient way.
The development of an efficient iterative solution scheme
indicates that the air-gap element approach is applicable to
technical models with a computational cost which is comparable to
the one of standard moving-band and sliding-surface techniques.



\printbibliography[heading=bibnumbered]

\nnsection{Authors name and Affiliation}

\noindent Herbert De Gersem and Thomas Weiland, Technische
Universit\"{a}t Darmstadt, Fachbereich 18 Elektrotechnik und
Informationstechnik, Institut f\"{u}r Theorie Elektromagnetischer
Felder (TEMF), Schlo{\ss}gartenstra{\ss}e 8, D-64289 Darmstadt,
Germany.

{\tt herbert.degersem@tu-darmstadt.de;
thomas.weiland@temf.tu-darmstadt.de}

\end{document}